 \documentclass[aip,jcp,reprint,amsmath,amssymb,showpacs]{revtex4-1}%
\usepackage{hyperref}
\usepackage{graphicx}
 \usepackage{amsmath}
\usepackage{graphicx}%
\usepackage{amsfonts}%
 \usepackage{amssymb}
\usepackage{color}
\usepackage{setspace}
\usepackage{textcomp}

\begin{document}

\title{From nanotubes to nanoholes: scaling of selectivity in uniformly charged nanopores through the Dukhin number for 1:1 electrolytes } 
\author{Zs\'ofia Sarkadi}
\author{D\'avid Fertig}
\author{Zolt\'an Hat\'o}
\author{M\'onika Valisk\'o}
\author{Dezs\H{o} Boda}\email[Author for correspondence:]{boda@almos.vein.hu}

\affiliation{$^{1}$Department of Physical Chemistry, University of Pannonia, P.O. Box 158, H-8201 Veszpr\'em, Hungary}
\date{\today}


\begin{abstract}
Scaling of the behavior of a nanodevice means that the device function (selectivity) is a unique smooth and monotonic function of a scaling parameter that is an appropriate combination of the system’s parameters. 
For the uniformly charged cylindrical nanopore studied here these parameters are the electrolyte concentration, $c$, voltage, $U$, the radius and the length of the nanopore, $R$ and $H$, and the surface charge density on the nanopore’s surface, $\sigma$. 
Due to the non-linear dependence of selectivites on these parameters, scaling can only be applied in certain limits.  
We show that the Dukhin number, $\mathrm{Du}=|\sigma|/eRc\sim |\sigma|\lambda_{\mathrm{D}}^{2}/eR$ ($\lambda_{\mathrm{D}}$ is the Debye length), is an appropriate scaling parameter in the nanotube limit ($H\rightarrow\infty$). 
Decreasing the length of the nanopore, namely, approaching the nanohole limit ($H\rightarrow 0$), an alternative scaling parameter has been obtained that contains the pore length and is called the modified Dukhin number: $\mathrm{mDu}\sim \mathrm{Du}\, H/\lambda_{\mathrm{D}}\sim |\sigma|\lambda_{\mathrm{D}}H/eR$.
We found that the reason of non-linearity is that the double layers accumulating at the pore wall in the radial dimension correlate with the double layers accumulating at the entrances of the pore near the membrane on the two sides. 
Our modeling study using the Local Equilibrium Monte Carlo method and the Poisson-Nernst-Planck theory provides concentration, flux, and selectivity profiles, that show whether the surface or the volume conduction dominates in a given region of the nanopore for a given combination of the variables. 
We propose that the inflection point of the scaling curve may be used to characterize the transition point between the surface and volume conductions.
\end{abstract}


\maketitle


\section{Introduction}
\label{sec:intro}

When the function of a device is determined by a few well-defined variables, $a_{1}, a_{2}, \dots$, it is often possible to group them into a composite parameter, $\xi$, that determines the device's behavior by itself.
This scaling parameter is a well-defined function of the independent variables: $\xi=\xi(a_{1}, a_{2}, \dots)$.
Let $F$ be the device function, an observable property of the device.
Scaling of the device function means that $F$ is a smooth unambiguous function of the scaling parameter: $F=f\left[ \xi(a_{1}, a_{2}, \dots) \right]$.

Nanopores are located in a membrane and connect two bath electrolytes.
They facilitate the controlled movement of ions from one side of the membrane to the other side.
Tunable input parameters of this device are the radius of the pore, $R$, the length of the pore (or, the width of the membrane), $H$, the voltage applied across the membrane, $U$, the surface charge on the wall of the nanopore, $\sigma$, ionic concentrations in the baths, $c$, and properties of the electrolyte, like ionic valences, $z_{i}$, and radii, $R_{i}$, for example ($i$ indexes the ionic species).
The measurable output parameters are the currents carried by the ionic species, $I_{i}$.

The structural properties of the nanopore (charge pattern and geometry) determine what is a practical choice for the device function.
In the case of a uniformly charged nanopore ($\sigma<0$ is constant) studied here (cation) selectivity defined as 
\begin{equation}
 S_{+}=\frac{I_{+}}{I_{+}+I_-} 
 \label{eq:sel}
\end{equation} 
is an appropriate device function and the in focus of our interest.
It is an unambiguous function of the currents and is well-measurable via the reversal potential.
For 1:1 electrolytes, if this number is $\approx 0.5$, the pore is non-selective, while if it is $1$, the pore is perfectly cation selective.
Note that $S_-=1-S_{+}$.

In a previous paper,~\cite{fertig_jpcc_2019} we introduced the scaling parameter $\xi=R/(\lambda \sqrt{z_{+}|z_-|})$ that was related to the $\lambda /R$ ratio, 
where $\lambda$ is a characteristic screening length of the electrolyte for which the Debye length is an obvious choice:
\begin{equation}
\lambda_{\mathrm{D}} = 
\left( \dfrac{ c e^{2}}{\epsilon_{0}\epsilon kT} \sum_{i} z_{i}^{2}\nu_{i} \right)^{-1/2} ,
\label{eq:lambdaD}
\end{equation} 
where $k$ is Boltzmann's constant, $T$ is the absolute temperature (it is $298.15$ K in this work), $e$ is the elementary charge, $\nu_{i}$ is the stoichiometric coefficient of ionic species $i$, $c$ is the salt concentration ($c_{i}=\nu_{i}c$ is the bath concentration of species $i$), $\epsilon$ is the dielectric constant of the solvent (it is $78.45$ in this work), and $\epsilon_0$ is the permittivity of vacuum.

The Debye length depends on the square root of the concentration, $\lambda_{\mathrm{D}}\sim 1{/}\sqrt{c}$, and characterizes the width of the double layer (DL) formed at the charged wall of the nanopore. 
Another choice \cite{fertig_jpcc_2019} for the screening length is the one obtained from the Mean Spherical Approximation (MSA)  \cite{blum_mp_1975,blum_jcp_1977,nonner_bj_2000} denoted by $\lambda_{\mathrm{MSA}}$. 
While the effect of the choice $\lambda = \lambda_{\mathrm{MSA}}$ was analyzed in our previous work,~\cite{fertig_jpcc_2019} we use the Debye length in this paper.
Some result using $\lambda_{\mathrm{MSA}}$ is shown in Fig.~ 1 of the Supporting Information (SI).

We showed \cite{fertig_jpcc_2019} that in a pore with a bipolar charge pattern (positive/negative) an obvious device function is the ratio of currents at forward- and reverse-biased values of voltage (rectification) and it scales with $\xi$.
In another paper, \cite{madai_pccp_2018} we showed that in a pore with a transistor-like charge pattern (positive/negative/positive) the ratio of currents in open and closed states (switching) scales with $\lambda_{\mathrm{D}}/R$ for a 1:1 electrolyte.

In these papers,~\cite{fertig_jpcc_2019,madai_pccp_2018} we studied nanopores with dimensions small enough that regions in the pore with very small ionic concentrations (depletion zones) form and determine device behavior.
The regions of a nanopore along the $z$-axis can be considered as resistors connected in series.
If the resistance of one segment is large due to the low concentration of an ionic species there (depletion zone), then the resistance is large for the whole pore for that ionic species.
Rectification of a bipolar nanopore, for example, is based on the fact that the depletion zones are deeper at one sign of the voltage than at the opposite sign.
A depletion zone for a given ionic species in a given region may form if that ionic species is the coion with respect to the surface charge in that region.

The formation of DLs, their overlap, and the resulting exclusion of coions are also key factors determining selectivity.
If the DL at the wall is wide compared to the pore radius ($\lambda_{\mathrm{D}} \gg R$), the DLs overlap and a bulk electrolyte is not formed around the centerline. 
The concentration of the coion is small not only close to the surface but also at the centerline.
The coions are depleted, counterions are in excess, and the surface conduction dominates.

If the width of the DL is small compared to the pore radius ($\lambda_{\mathrm{D}} \ll R$), the DL is restricted to the region close to the surface and the region along the centerline contains enough coions so they can carry current in that region.
In the region where both coions and counterions are present, volume conduction dominates.

Between the two limiting cases both surface and volume conductions are present and our negative pore is not perfectly cation selective ($S_{+}$ is in between $0.5$ and $1$ for the 1:1 electrolyte considered here).

The calculations reported in Ref.~\onlinecite{fertig_jpcc_2019} were performed for a fixed value of the surface charge ($\sigma = \pm 1$ $e$/nm$^{2}$) and of the pore length ($H=6$ nm), so the $\xi$ scaling parameter did not contain $\sigma$ and $H$.
Here, we pursue a scaling parameter that also includes $\sigma$ and $H$ in addition to $R$ and $c$.

\begin{figure}[t!]
\begin{center}
\includegraphics*[width=0.49\textwidth]{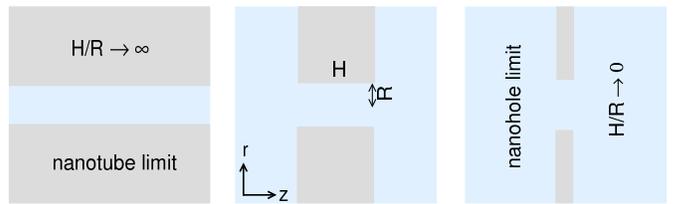}
\end{center}
\caption{Pore geometries from the nanotube limit, $H/R\rightarrow\infty$, to the nanohole limit, $H/R\rightarrow 0$.
}
\label{fig1}
\end{figure}

Changing pore length $H$ at a fixed $R$ means changing the aspect ratio of the pore, $H/R$ (see Fig.~\ref{fig1}).
We will start from the infinitely long pore, where $H$ and $U$ do not appear as variables, and decrease the pore length approaching small $H/R$ values.
The $H/R\rightarrow \infty$ and $H/R\rightarrow 0$ limits are called the nanotube and nanohole limits, respectively.
While the nanotube limit is well known for example in PET nanopores~\cite{Siwy_2004}, the nanohole limit can also achieved by increasing the radius of the pore or by using a very thin membrane, graphene, for example~\cite{garaj_n_2010,garaj_pnas_2013}.
We show that the DLs formed at the entrances of the pore near the surfaces of the membrane have a serious effect on the behavior of the pore in general, and on scaling specifically.

\section{Model of the nanopore}
\label{sec:model}

A cylindrical nanopore of length $H$ and radius $R$ spans a membrane that separates two baths.
The cylindrical wall of the nanopore and the flat parallel walls confining the membrane are assumed to be hard. 
The $z$ dimension is the one perpendicular to the membrane along the pore.
Because the system has rotational symmetry about the $z$ axis, the other relevant coordinate is the radial one, $r$, which represents the distance from the $z$ axis (Fig.~\ref{fig1}).

The electrolyte is modeled in the implicit solvent framework, namely, the interaction potential between two hard-sphere ions is defined by Coulomb's law in a dielectric background when the ions are not overlapped:
\begin{equation}
 u_{ij}(r')=
 \left\lbrace 
\begin{array}{ll}
\infty & \quad \mathrm{if} \quad r'<R_{i}+R_{j} \\
 \dfrac{1}{4\pi\epsilon_{0}\epsilon} \dfrac{z_{i}z_{j}e^{2}}{r'} & \quad \mathrm{if} \quad r' \geq R_{i}+R_{j}\\
\end{array}
\right. 
\label{eq:uij}
\end{equation}
where $R_i$ is the radius of ionic species $i$, and $r'$ is the distance between two ions. 
Here, we consider only 1:1 electrolytes, namely, $z_{+}=1$ and $z_-=-1$.
For the ionic radii, we use $R_{+}=R_-=0.15$ nm.
A uniform negative surface charge $\sigma$ is placed on the wall of the nanopore.

The description of the methods with which we study this model is included in the Appendix.
Both methods are based on the Nernst-Planck (NP) transport equation~\cite{nernst_zpc_1888,planck_apc_1890}:
\begin{equation}
 \mathbf{j}_{i}(\mathbf{r})= -\frac{1}{kT} D_{i}(\mathbf{r})c_{i}(\mathbf{r})\nabla \mu_{i}(\mathbf{r}),
 \label{eq:np}
\end{equation} 
where $\mathbf{j}_{i}(\mathbf{r})$, $D_{i}(\mathbf{r})$, $c_{i}(\mathbf{r})$, and $\mu_{i}(\mathbf{r})$ are the flux density, the diffusion coefficient profile, the concentration profile, and the electrochemical potential profile of ionic species $i$, respectively.
To make use of this equation, we need a relation between the concentration profile, $c_{i}(\mathbf{r})$, and the electrochemical potential profile, $\mu_{i}(\mathbf{r})$.
Note that we ignore the motion of the solvent here due to the relatively small pore radii and voltages.

In one method, we relate the concentration profile, $c_{i}(\mathbf{r})$, to the electrochemical potential profile, $\mu_{i}(\mathbf{r})$, with the Poisson-Boltzmann (PB) theory.
This continuum theory is known as the Poisson-Nernst-Planck (PNP) theory.

The other method is based on a Monte Carlo (MC) technique that is an adaptation of the Grand Canonical Monte Carlo (GCMC) method to a non-equilibrium situation, where $\mu_{i}(\mathbf{r})$ is not constant system-wide: the system is not in global equilibrium, only in local equilibrium.
The method is called the Local Equilibrium Monte Carlo (LEMC) technique, while it is called NP+LEMC when we couple it to the NP equation.


We also study the nanotube limit ($H\rightarrow \infty$), where the length of the pore is much larger than its radius.
This limit can be estimated from simple equilibrium calculations because in an infinitely long nanotube the electric field (the driving force of the ion transport) is constant, so the flux density is proportional to the concentration, $j_{i}(r)\sim c_{i}(r)$, due to the NP equation (Eq.~\ref{eq:np}).
Currents, therefore, are proportional to the cross-sectional integrals of the concentrations:
\begin{equation}
 I_{i}=z_{i}e\int_{A}j_{i}(r)\mathrm{d}a \sim \int_{A}c_{i}(r) \mathrm{d}a.
\end{equation} 
The concentration profile can be calculated from equilibrium simulation that corresponds to the zero-voltage limit (slope conductance).

The $H\rightarrow \infty$ limit of NP+LEMC can be computed from equilibrium GCMC simulations, where the tube is infinite in the sense that a periodic boundary condition is applied in the $z$ direction.
Insertions/deletions of neutral ion pairs (a cation and an anion) connect the tube to the bath of a fixed chemical potential $(\mu_{+}+\mu_{-})/2$ that corresponds to the prescribed salt concentration.~\cite{malasics_jcp_2008,malasics_jcp_2010}
In addition to the concentration profiles, $c_{i}(z,r)$, the simulations directly provide the average numbers of ionic species in the tube, $\langle N_{i}\rangle$, from which selectivity follows.

The equilibrium and $H\rightarrow \infty$ limit of PNP can be computed by solving the PB equation
\begin{equation}
 \frac{1}{r} \frac{\mathrm{d}}{\mathrm{d} r} \left( r \frac{\mathrm{d} \Phi(r)}{\mathrm{d} r} \right) = \frac{1}{\lambda_{\mathrm{D}}^{2}} \Phi(r)
 \label{eq:pb}
\end{equation} 
with the boundary conditions that $\mathrm{d}\Phi(r=0)/\mathrm{d}r=0$ and $\Phi(R)=\Phi_{0}$, where $\Phi_{0}$ is a prescribed potential at the wall. 
The surface charge is computed from the boundary condition of $\mathrm{d}\Phi(r)/\mathrm{d}r$ at $r=R$.
The concentrations are obtained from $c_{i}(r)=c_{i}\exp\left( -z_{i}e\Phi(r)/kT \right)$, where $c_{i}$ is the bulk concentration of ionic species $i$.
While analytic solutions are available for the limits of dominant bulk conduction ($R\ll \lambda_{\mathrm{D}}$) and dominant surface conduction ($\lambda_{\mathrm{D}} \ll R$),~\cite{levine_jcis_1975,balme_sr_2015,uematsu_jpcb_2018,green_jcp_2021} we are bound to solving the PB equation numerically in between.

\section{Results}
\label{sec:results}

Before we study scaling, we briefly analyze how the system behaves in the radial and axial directions, namely, how the concentration profiles behave in these dimensions.


\subsection{Radial dimension: double layer formation/overlap near the pore wall}

\begin{figure}[t!]
\begin{center}
\includegraphics*[width=0.49\textwidth]{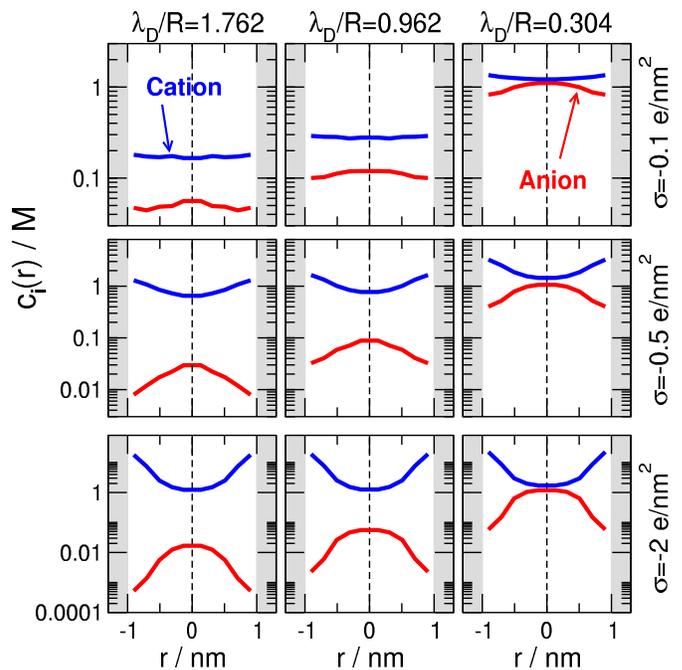}
\end{center}
\caption{Radial concentration profiles (obtained by averaging over the pore in the $z$ direction) for various combinations of $\sigma$ and $\lambda_{\mathrm{D}}/R$ for fixed $H=6$ nm, $R=1$ nm, and $U=200$ mV.
Different columns have different values of $\lambda_{\mathrm{D}}/R$, while different rows have different values of $\sigma$.
The concentrations that correspond to the $\lambda_{\mathrm{D}}/R =1.762$, $0.962$, and $0.304$ values are $c=0.03$, $0.1$, and $1$ M, respectively.
Blue and red curves refer to the cations and the anions, respectively.
The curves are mirrored to $r=0$ for better visualization, while strictly speaking $r{\geq} 0$.
}
\label{fig2}
\end{figure}

Figure \ref{fig2} illustrates the effect of the two relevant parameters ($\sigma$ and $\lambda_{\mathrm{D}}/R$)  for fixed values of $R$, $H$, and $U$. 
The surface charge produces the separation of counterions and coions (DL formation), while the $\lambda_{\mathrm{D}}/R$ parameter accounts for the degree of overlap of the DL.
The figure shows radial concentration profiles to depict the interplay of these two effects. 

The columns correspond to different $\lambda_{\mathrm{D}}/R$ values (a larger $\lambda_{\mathrm{D}}/R$ value means stronger DL overlap), while the rows correspond to different surface charges (a larger $|\sigma|$ value results in a stronger separation of cation and anion profiles).

\begin{figure*}[t!]
\begin{center}
\includegraphics*[height=0.35\textwidth]{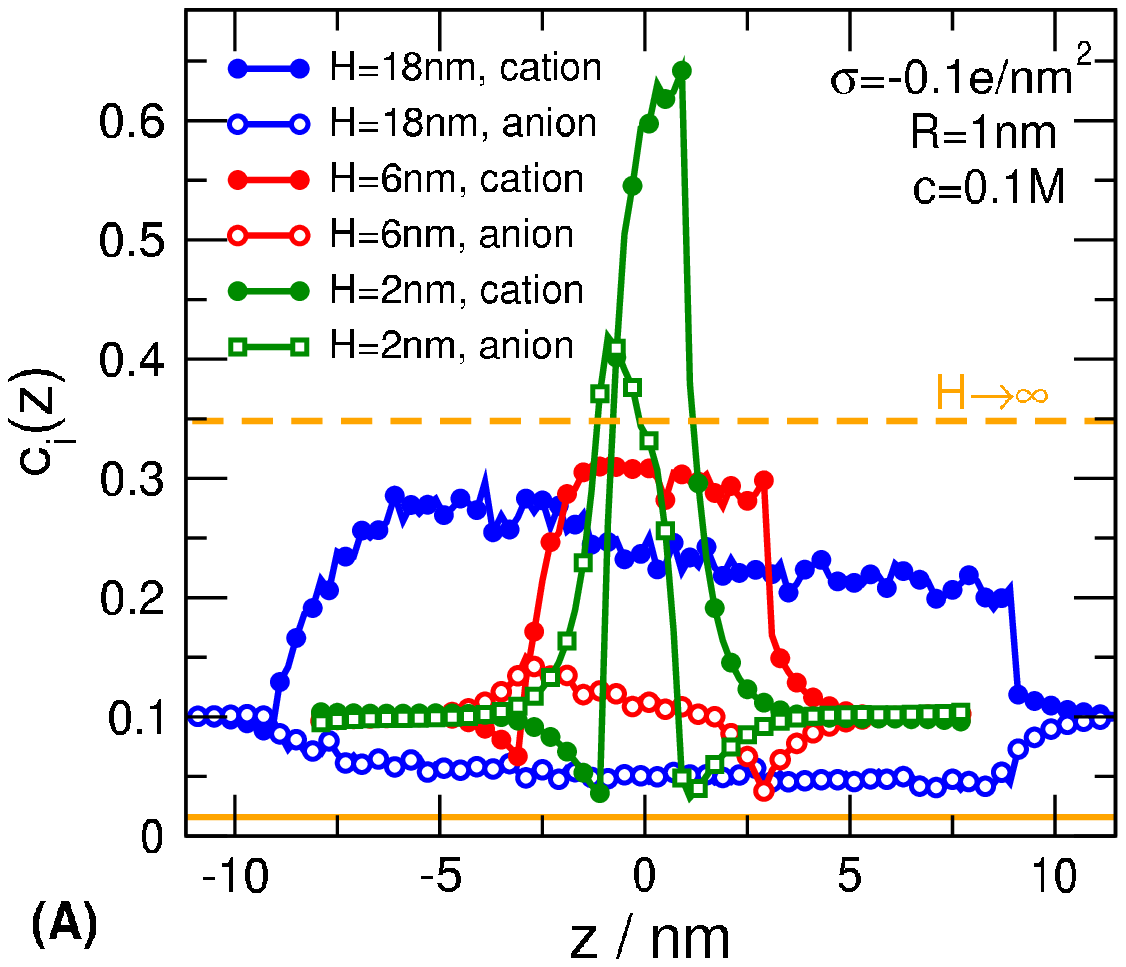} \hspace{0.5cm}
\includegraphics*[height=0.35\textwidth]{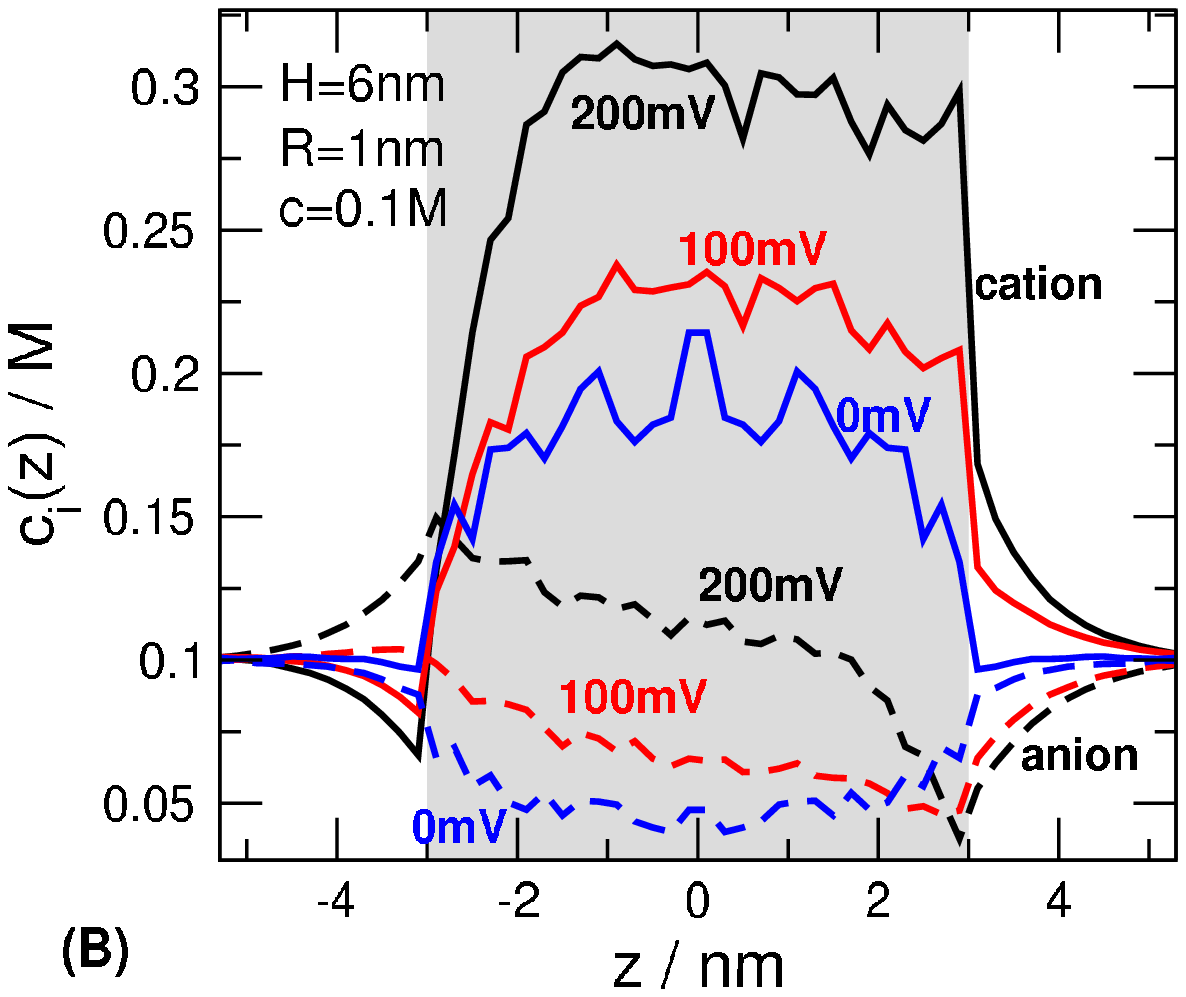}
\end{center}
\caption{(A) Axial concentration profiles (obtained by averaging over the pore in the $r$ direction) for various pore lengths ($H=2$, $6$, $18$ nm)  for $\sigma=-0.1$ $e$/nm$^{2}$, $R=1$ nm, and $c=0.1$ M as obtained from NP+LEMC simulations using $200$ mV voltage. 
Filled and open symbols refer to cations and anions, respectively.
The dashed and solid horizontal orange lines indicate the cation and anion profiles for the $H\rightarrow\infty$ limit.
(B) Axial concentration profiles for various voltages for pore length $H=6$ nm. Solid and dashed lines refer to cations and anions, respectively.
The shaded grey area indicates the pore region.
}
\label{fig3}
\end{figure*} 

As we go from left to right (decreasing $\lambda_{\mathrm{D}}/R$), the gap between the cation and anion profiles around the centerline $r\approx0$ decreases (note the log scale).
As we go from top to bottom (increasing $|\sigma|$), the gap between the cation and anion profiles near the wall ($r\approx R$) increases.

This implies that the two parameters have two relatively distinct effects.
The $\lambda_{\mathrm{D}}/R$ parameter rather determines the behavior in the middle of the pore, while $\sigma$ rather determines the behavior in the DL.
It is common to express this distinction in terms of volume (or bulk) and surface conductions.
The DL region (if exists) is responsible for the surface conduction, while the bulk region in the middle (if exists) is responsible for the volume conduction.
With $\lambda_{\mathrm{D}}/R$ and $\sigma$, therefore, we have two parameters with which we can tune the weights of the volume and surface conductions in the total conduction.

These two effects, however, separate only in the nanotube limit ($H\rightarrow\infty$) that physically can be corresponded to a finite but long nanopore, where the influence of the DLs at the entrances of the nanopore is negligible in the middle of the nanopore.
Voltage can also modify the DLs at the membrane's two sides: it creates DLs of opposite signs on the two sides that, in turn, also modify the ionic distributions inside the pore.
If the pore is short, therefore, these effects must be taken into account.


\subsection{Axial dimension: the effect of pore entrances inside the pore}

Considering the axial dimension, our main question is how the length of the pore influences the concentration profiles in the pore, and, consequently, selectivity.
Fig.~\ref{fig3}A shows the profiles for different values of the pore length.
The cation concentration is larger in the $H\rightarrow\infty$ limiting case than in the $H=18$ nm case (the anion concentration, at the same time, is smaller).
This is because charge neutrality is enforced in the GCMC simulations for $H\rightarrow\infty$, while the finite pore does not have to be charge neutral.

The pore charge of a short pore is only partly neutralized by the excess counterions inside the pore; it is partly neutralized by the excess cations accumulated at the entrances of the pore in the DLs near the membrane.
This accumulation is larger if $\sigma$ is larger. 
Because the ions are charged hard spheres, volume exclusion works against ions entering the pore and ion accumulation outside (where there is more space) is advantageous and minimizes free energy. 

This counterion accumulation is shown in Fig.~\ref{fig3}B by the curves for $U=0$ mV.
When voltage is increased, an excess cation accumulation can be observed on the right hand side, while an excess anion accumulation on the left hand side of the pore.
This pair of charge accumulations of opposite signs creates a ``dipole field'' and arises because the electric field of the ions produces a counterfield against the applied field.
More details about the origin of this ``dipole field'' can be found in the Appendix (Fig.~\ref{figapp}), where it is explained that a larger charge accumulation is obtained both for shorter pores and for larger voltages.
This is clearly seen from the concentration profiles of Fig.~\ref{fig3} as well.

As the pore length is decreased further from $H=18$ nm, the concentration profiles of both cations and anions elevate in the pore, because the DLs contain more charge in the case of short pores and the stronger counterion accumulation has a larger effect inside the pore.
This modifies the imbalance of cations and anions in the pore for different $H$ values results in varying selectivity.
Similarly, increasing  voltage increases the degree of charge separation between the two sides of the membrane (Fig.~\ref{fig3}B).
By the same mechanism, the changed DL structure on the two sides of the membrane changes the electric field inside the pore, and, consequently, ionic distributions and selectivity. 

In both panels of Fig.~\ref{fig3}, the axial profiles have been obtained by normalizing with the cross section of the cell.
Thus, the concentration profiles characterize the density of ions at a given $z$, not the quantity of ions.
The quantity of ionic charge in the DLs outside the pore, therefore, is much larger than implied by this figure.
Keeping this in mind, however, the figure shows the trends and describes the $H$ dependence (panel A) and the $U$ dependence (panel B) of the ionic profiles.

If we want to assess how the pore entrances influence the interior of the pore, we can use the relation of the screening length to the pore length, $\lambda_{\mathrm{D}}/H$, to quantify this effect.
The charge accumulation at the entrance creates a double layer in the axial dimension that ``penetrates'' the pore. 
The degree of penetration and whether it reaches the middle of the pore can be characterized, at least, as a first approximation, with the ratio $\lambda_{\mathrm{D}} /H$. 
If $\lambda_{\mathrm{D}} /H>0.5$, the DLs extending axially from the two sides overlap in the middle, a phenomenon we call ``bridging''.

From these analyses, we can see that the picture gets complicated as we go from the infinite pore to shorter pores and larger voltages.
Every parameter has an effect that also influences the effects of other parameters.
In the following, we try to cut a path through this jungle and provide a systematic analysis of the effects of all parameters.
Because the nanotube limit is the simplest one, it is reasonable to start our analysis with that case.


\subsection{The Dukhin number is the scaling parameter in the nanotube limit}

Because there is a monotonic relationship between ionic selectivity and the ratio of surface and volume conductances that are present in the nanopore at the same time, the Dukhin number \cite{bazant_pre_2004,chu_pre_2006,bocquet_chemsocrev_2010} defined as
\begin{equation}
 \mathrm{Du} \equiv \frac{|\sigma|}{eRc} 
 \label{eq:dukhin}
\end{equation} 
offers itself to be the scaling parameter we are after. 
This definition of Du contains exactly those variables that we want to see in our scaling parameter, $\sigma$, $R$, and $c$, although $H$ is missing. 

Du was originally introduced by Bikerman \cite{bikerman_1940} to characterize the ratio of the surface and volume conductances focusing on electrokinetic phenomena.
Later, Dukhin adopted the idea (see Ref.~\onlinecite{dukhin_advcollsci_1993} and references therein) to study electrophoretic phenomena.
Although the name Bikerman number also occurs in the literature, Lyklema introduced the name Dukhin number to salute Dukhin.~\cite{lyklema_book_1995}
The Dukhin number (and a characteristic length called Dukhin length, $l_{\mathrm{Du}}=\mathrm{Du}\, R$) was used in several modeling studies describing nanopores, more specifically, when volume and surface transport processes competed inside the pore.~\cite{bazant_pre_2004,chu_pre_2006,khair_jfm_2008,das_langmuir_2010,bocquet_chemsocrev_2010,zangle_csr_2010,lee_nanolett_2012,yeh_ijc_2014,ma_acssens_2017,xiong_scc_2019,poggioli_jpcb_2019,dalcengio_jcp_2019,kavokine_annualrev_2020,noh_acsnano_2020}

The definition in Eq.~\ref{eq:dukhin} is in agreement with the traditional definition of Bikerman because it can be computed from the ratio of the surface excess of the cations, $|\sigma| 2\pi R H$ assuming perfect exclusion of the anions, and the number of charge carriers assuming a bulk electrolyte in the pore, $2c R^{2}\pi H$ (the factor $2$ is needed because both cations and anions carry current). 

\begin{figure}[t]
\begin{center}
\includegraphics*[width=0.4\textwidth]{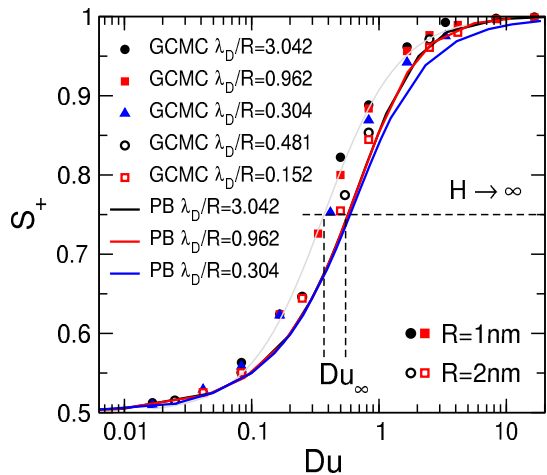}
\end{center}
\caption{Selectivity curves obtained by scanning the surface charge, $\sigma$, for an infinitely long nanopore ($H\rightarrow\infty$, nanotube limit) as functions of the Dukhin number (Eqs.~\ref{eq:dukhin} and \ref{eq:du1}). The curves have been obtained by numerically solving the PB equation (Eq.~\ref{eq:pb}) for different values of $\lambda_{\mathrm{D}}/R$. The indicated values correspond to 1:1 electrolytes of concentrations $c=0.01$, $0.1$, and $1$ M for pore radius $R=1$ nm.
Filled and open symbols indicate the results of GCMC simulations for hard-sphere ions (Eq.~\ref{eq:uij}) for these concentrations. The filled and open symbols correspond to GCMC simulations for $R=1$ and $2$ nm, respectively. 
The inflection points of the PB and GCMC curves are indicated as $\mathrm{Du}_{\infty}$. 
The estimated  values are $\mathrm{Du}_{\infty}\approx 0.37$ and $0.546$ for GCMC and PB, respectively.
}
\label{fig5}
\end{figure} 

For a 1:1 electrolyte, the Debye length can be written as $\lambda_{\mathrm{D}}^{2}=1/(8\pi l_{\mathrm{B}} c)=1/(l_{\mathrm{B}}^{*}c)$, where $l_{\mathrm{B}}^{*}=8\pi l_{\mathrm{B}}$ and $l_{\mathrm{B}}=e^{2}/4\pi \epsilon_0 \epsilon kT$ is the Bjerrum length.
The Dukhin number then can be rewritten in the form
\begin{equation}
 \mathrm{Du}=\frac{|\sigma| l_{\mathrm{B}}^{*}\lambda_{\mathrm{D}}^{2}}{eR}
 \label{eq:du1}
\end{equation} 
that makes it possible to relate quantities with the dimensions of distance to each other.
We may also use other variations for the screening length that may depend on electrolyte properties ($z_{i}$ and $R_{i}$) such as the MSA screening length \cite{fertig_jpcc_2019} or on confinement~\cite{levy_jcis_2020}.

Figure \ref{fig5} shows the selectivity (Eq.~\ref{eq:sel}) as a function of $\mathrm{Du}$ (log scale for Du). 
The figure includes results from PB calculations (lines) and GCMC simulations (symbols). 
In the case of PB, the Debye length and pore radius are coupled, so the curves for different $\lambda_{\mathrm{D}}/R$ values collapse onto one single curve. 
The deviations for the small $\lambda_{\mathrm{D}}/R$ value are due to  to the imprecision of the numerical methods applied).
This problem only appeared at higher concentration values as the slope of the different functions ($\Phi(r), c_{i}(r)$) became larger.
The figure also shows the Du values of the inflection points of the curves denoted as $\mathrm{Du}_{\infty}$.
This value is estimated as $\mathrm{Du}_{\infty}\approx 0.546$ for PB.

The inflection point belongs to selectivity $S_{+}\approx 0.75$ and have a special role in our analysis. 
Fig.~\ref{fig5} and later figures show that the selectivity curves can be fitted with sigmoid curves
\begin{equation}
 S_{+}(x) = 0.5 + \dfrac{0.5}{1+e^{-k(x-x_{0})}},
 \label{eq:sigmoid}
\end{equation} 
where $x=\lg(\mathrm{Du})$ and $x_{0}$ defines the inflection point ($\lg$ denotes logarithm of base $10$).
The shape of the curves is very similar (the parameter $k$ is in the same ballpark between $2.5$ and $4$), so the major difference is that the selectivity curves are shifted along with the $\lg(\mathrm{Du})$ axis as $H$, $U$ or any other parameter is changed.

This way, we reduced the problem to the examination of the shift of the inflection point denoted as $\mathrm{Du}_{0}$.
For the nanotube, therefore, $\mathrm{Du}_{\infty}=\lim_{H\rightarrow \infty} \mathrm{Du}_{0}(H)$.
Because we have a good scaling for $H\rightarrow \infty$ , the $\mathrm{Du}_{\infty}$ values have a special role: we can relate the cases for finite pores to this limit.
If we divide $\mathrm{Du}$ with $\mathrm{Du}_{0}$, we shift the selectivity curve into $\mathrm{Du}^{*}=\mathrm{Du}/\mathrm{Du}_{0}=1$, where 
\begin{equation*}
 \mathrm{Du}^{*}=\frac{\mathrm{Du}}{\mathrm{Du}_{0}}
\end{equation*} 
is a rescaled Dukhin number.

\begin{figure*}[t!]
\begin{center}
\includegraphics*[width=0.7\textwidth]{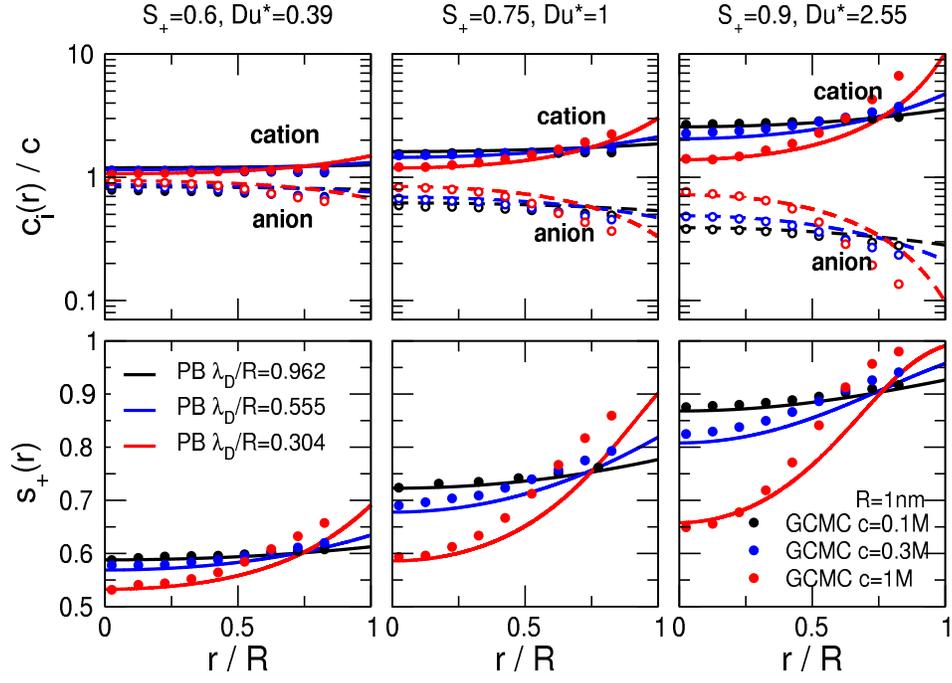}
\end{center}
\caption{Radial concentration profiles (top row) and selectivity profiles (bottom row) for different values of $\lambda_{\mathrm{D}}/R$ as obtained from PB calculations (lines) and equilibrium GCMC simulations (symbols).
The concentrations that correspond to the indicated $\lambda_{\mathrm{D}}/R$ values are $c=0.1$, $0.3$, and $1$ M for $R=1$ nm. The symbols have been obtained for these state parameters.
The columns of the figure correspond to fixed selectivities and to the corresponding $\mathrm{Du}^{*}$ numbers.
}
\label{fig6}
\end{figure*} 

The $\mathrm{Du}_{0}$ inflection point, of course, depends on every parameter and the model (NP+LEMC or PNP).
Our purpose is to map this dependence in a way that is useful and consumable for the reader (the number of variables is large).

For GCMC, for example, $\mathrm{Du}_{\infty}\approx 0.37$ as shown by Fig.~\ref{fig5}.
The deviation between PB and GCMC is twofold.
First, the ions are point-like in PB, while they have finite size in the GCMC simulation (Eq.~\ref{eq:uij}).
Second, the GCMC (and LEMC) simulations include all the electrostatic correlations that are beyond the mean-field (BMF) treatment of PB (and PNP).
For the 1:1 electrolyte studied here, the electrostatic BMF correlations are relatively weak, so the main source of the deviation between PB and GCMC is the finite size of particles.

This statement is supported by Fig.~\ref{fig5} that shows GCMC results for various $\lambda_{\mathrm{D}}/R$ ratios obtained from simulations for $R=1$ and $2$ nm using various concentrations.
The points obtained for $R=2$ nm (open symbols) are closer to the PB curves because the finite size of ions has less importance in the wide pore.

\begin{figure*}[t!]
\begin{center}
\includegraphics*[width=0.95\textwidth]{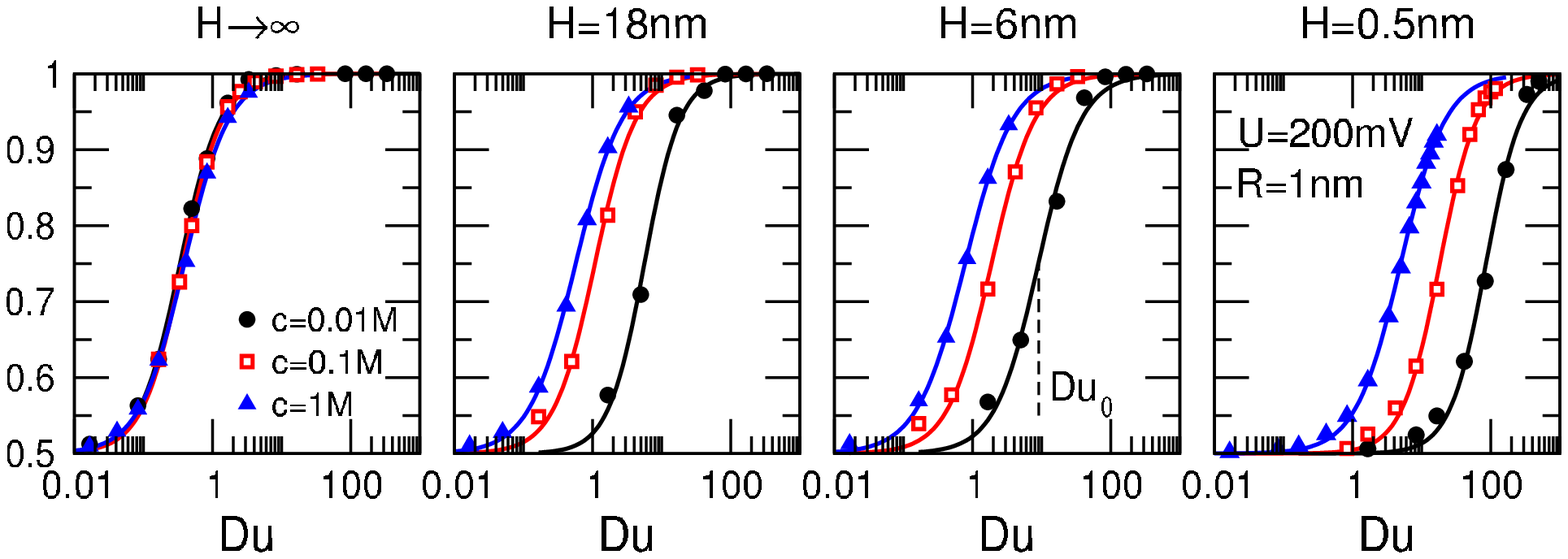}
\end{center}
\caption{Selectivity curves obtained by scanning the surface charge, $\sigma$, for nanopores of decreasing lengths from left to right (from $H\rightarrow\infty$ to $H=0.5$ nm) as functions of the Dukhin number (Eqs.~\ref{eq:dukhin} and \ref{eq:du1}). 
The radius of the pore and voltage are fixed ($R=1$ nm and $U=200$ mV).
The symbols have been obtained from NP+LEMC simulations (GCMC for $H\rightarrow\infty$) for different concentrations, while the lines are fitted sigmoid curves (Eq.~\ref{eq:sigmoid}). 
$\mathrm{Du}_{0}$ denotes the inflection points of the curves.
}
\label{fig8}
\end{figure*}

The difference between PB and GCMC can also be seen in Fig.~\ref{fig6} that shows concentration profiles (top row) and selectivity profiles (bottom row) defined as 
\begin{equation}
 s_{+}(r) = \dfrac{j_{+}(r)}{j_{+}(r)+j_-(r)} \approx \dfrac{c_{+}(r)}{c_{+}(r)+c_-(r)},
\end{equation} 
where $j_{+}(r)$ and $j_-(r)$ are the $z$-components of the flux profiles for the cation and the anion, respectively, averaged over the pore in the $z$ dimension (from $-H/2$ to $H/2$).
It is a quantity that depends on $r$ and characterizes to what degree a region of the pore at a distance $r$ from the $z$ axis contributes to the ``global'' selectivity, $S_{+}$.
Note that the average of $s_{+}(r)$ is not equal to $S_{+}$, but we can draw conclusions from the shapes of the curves nevertheless.

The columns of Fig.~\ref{fig6} refer to different selectivities corresponding to specific scaled Dukhin numbers, $\mathrm{Du}^{*}$. 
These different $\mathrm{Du}^{*}$ values correspond to different combinations of $\sigma$ and $\lambda_{\mathrm{D}}/R$.
If $\lambda_{\mathrm{D}}/R$ is small (red curves), for example, larger $\sigma$ values are needed to achieve the same selectivity.
Red curves for $s_{+}(r)$ rise to a higher value at the wall due to the increased $\sigma$, but they decrease (the $c_{+}(r)$ and $c_{-}(r)$ profiles get closer to each other) in the centerline of the pore resulting in the same global selectivity, $S_{+}$, as, for example, the black curves.
In the case of the black curves, $\lambda_{\mathrm{D}}/R$ is large, so the DL overlap is strong.
As a consequence, a smaller surface charge is sufficient to get the same selectivity. 

The difference between the PB and GCMC curves is apparent.
Because the centers of the ions are confined into a cylinder with the radius $R-R_{i}$ ($R_{i}=0.15$ nm), the ionic profiles do not extend to the wall.
If we replot the figure by normalizing $r$ with $R-R_{i}$ instead of $R$, we obtain a much better agreement (see Fig.~2 of the SI).
The $R-R_{i}$ radius can be considered an ``effective'' radius of the pore that the centers of the hard-sphere ions can reach.

\subsection{Scaling for finite pores}

If we start decreasing the pore length (voltage is $U=200$ mV), the DLs at the entrances of the pore ``penetrate'' the pore and modify the ionic distributions inside the pore as shown in Fig.~\ref{fig3}A (``bridging'').
This modifies the mutual effects of the parameters on selectivity, so also the shift of the $S_{+}(\mathrm{Du})$ function with respect to $\mathrm{Du}_{\infty}$.
This is shown in Fig.~\ref{fig8}, where the pore length decreases from left to right and different colors in a panel belong to different concentrations.

The curves are shifted to the right, so shorter pores require larger surface charges to produce the same selectivity.
The curves for lower concentrations (at a given $H$) are also shifted to the right so electrolytes with more extended diffuse layers enhance the effect of the small length of the pore and also require larger surface charges to produce the same selectivity.

The main conclusion of this figure is that Du is not an appropriate scaling parameter for finite pores.

\begin{figure*}[t!]
\begin{center}
\includegraphics*[height=0.38\textwidth]{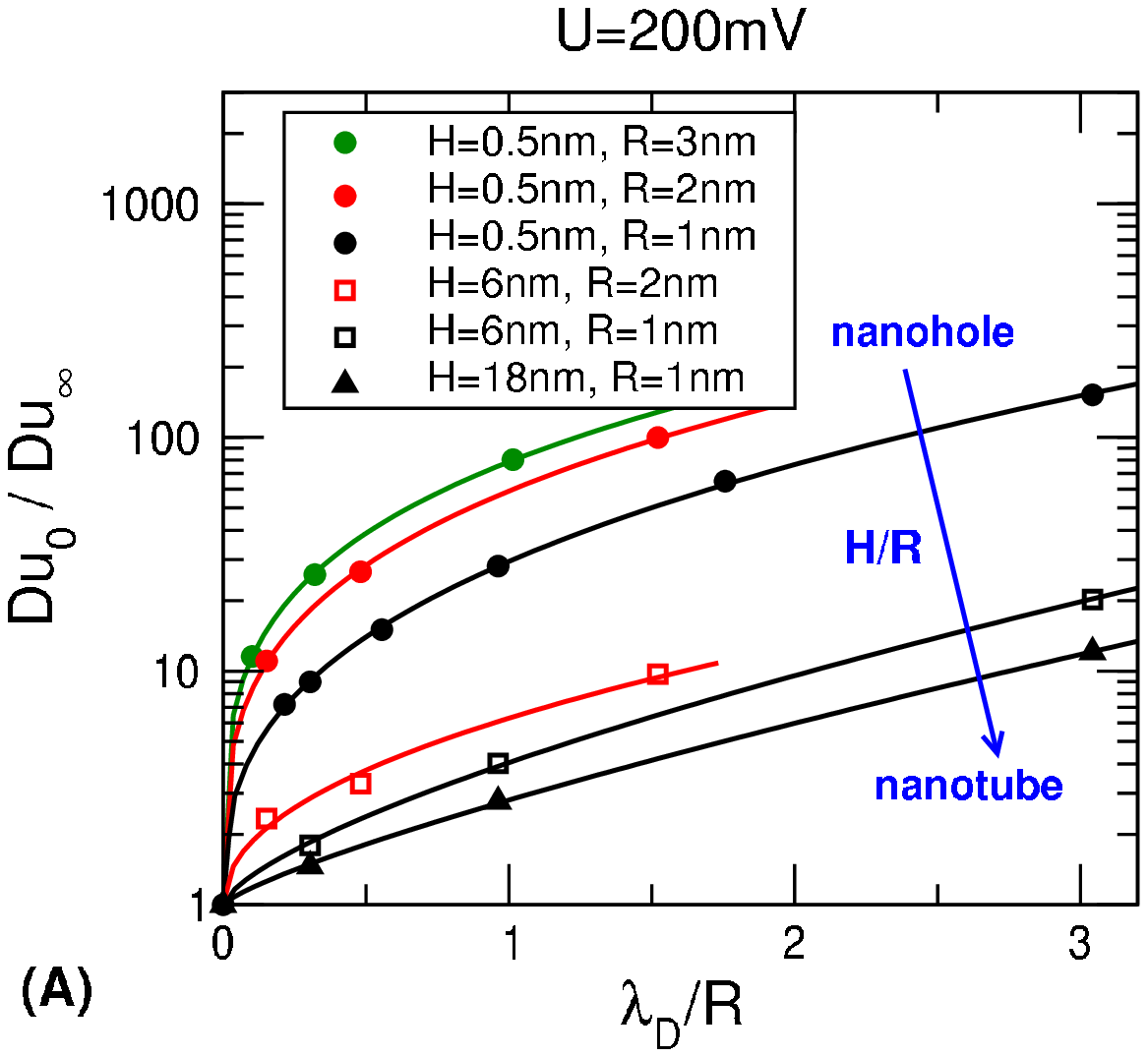} \hspace{0.5cm}
\includegraphics*[height=0.38\textwidth]{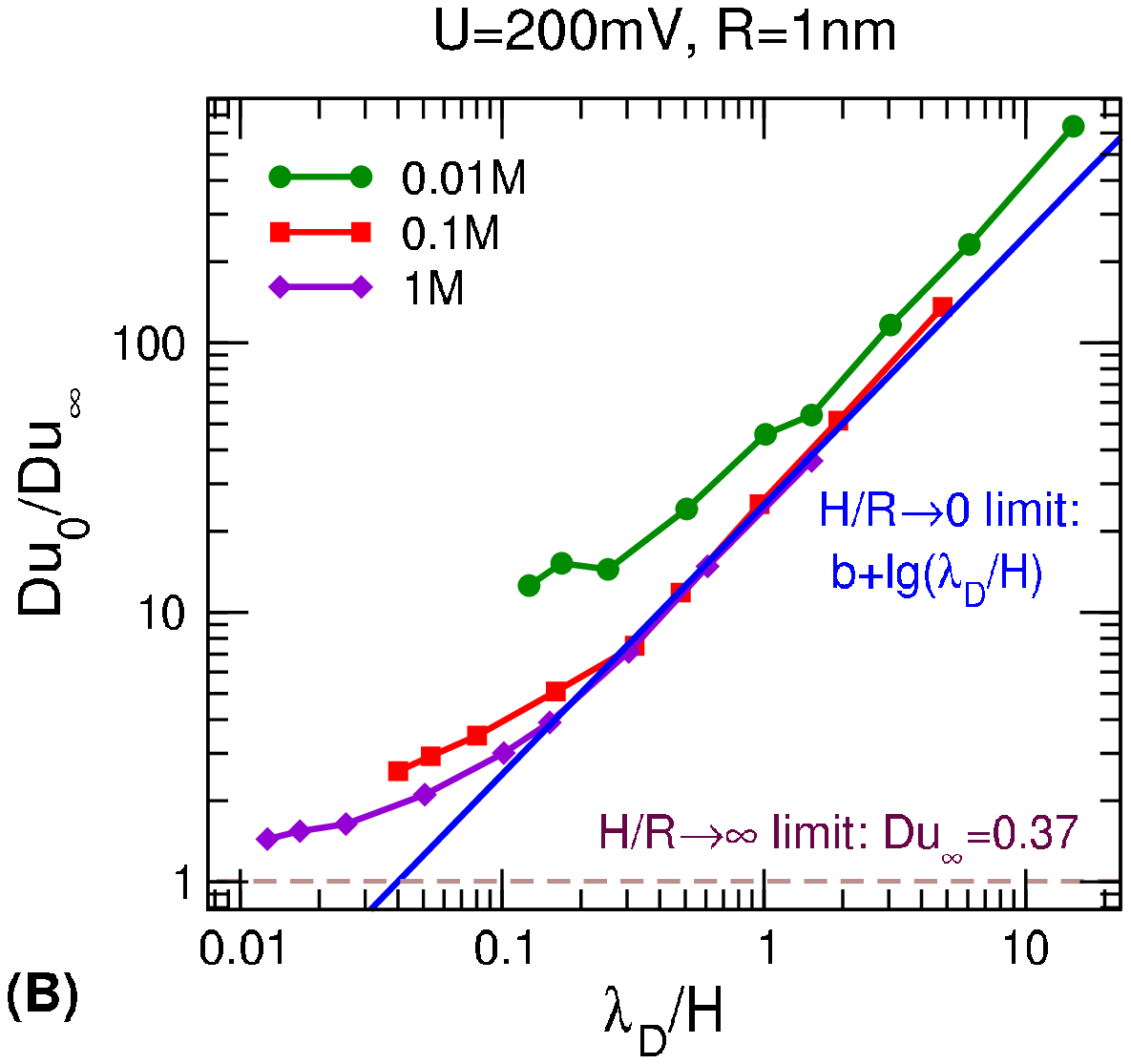}
\end{center}
\caption{The inflection point of the $S_{+}(\mathrm{Du})$ curves, $\mathrm{Du}_{0}$, normalized by $\mathrm{Du}_{\infty}$ as a function of (A) $\lambda_{\mathrm{D}}/R$ and (B) $\lambda_{\mathrm{D}}/H$ as obtained from NP+LEMC simulations for $U=200$ mV. Panel A shows results for increasing aspect ratio, $H/R$, (indicated by the arrow) for different pairs of $R$ and $H$. The lines are power function fits. Panel B shows results for fixed $R$ and different concentrations. Curves for different concentrations have been obtained by changing $H$. The blue line shows a linear fit to points of small $c$ and small $H$ (Eq.~\ref{eq:linfit}). 
}
\label{fig9}
\end{figure*} 

Fig.~\ref{fig9} shows the inflection point, $\mathrm{Du}_{0}$ normalized by the $H\rightarrow\infty$ value, $\mathrm{Du}_{\infty}$ as a function of $\lambda_{\mathrm{D}}/R$ (Fig.~\ref{fig9}A) and  $\lambda_{\mathrm{D}}/H$ (Fig.~\ref{fig9}B).
So, the limiting value $\mathrm{Du}_{0}/\mathrm{Du}_{\infty}=1$ corresponds to the nanotube limit.
Fig.~\ref{fig9}A shows the data as functions of $\lambda_{\mathrm{D}}/R$ for different aspect ratios $H/R$.
With increasing $\lambda_{\mathrm{D}}/R$, we have wider DLs for a fixed $R$, so these wider DLs extend into the pore changing nanotube behavior and shifting the inflection point.
With increasing $H/R$ (indicated by the arrow in the figure), we approach the nanotube limit, so the values of $\mathrm{Du}_{0}/\mathrm{Du}_{\infty}$ decrease for a fixed $\lambda_{\mathrm{D}}/R$ approaching the $\mathrm{Du}_{0}/\mathrm{Du}_{\infty}=1$ limit.
The lines are fitted power functions.
Unfortunately, we were not able to find any pattern in the exponent.

If we plot the results as functions of $\lambda_{\mathrm{D}}/H$, on the other hand, we can characterize the two limiting cases quantitatively (Fig.~\ref{fig9}B).
For a fixed pore radius $R=1$ nm, the curves of a given color correspond to a given concentration.
As $\lambda_{\mathrm{D}}/H$ decreases, we approach the nanotube limit.
Either the pore becomes long enough for a given electrolyte so that the pore entrances do not disturb the ions' behavior in the middle of the pore, or the electrolyte becomes more concentrated so that the DLs extend less into the pore.
The system approaches this limiting case with different trends depending on concentration.
At large concentration ($c=1$ M, purple curve) the curve approaches the $\mathrm{Du}_{0}/\mathrm{Du}_{\infty}=1$ limit faster as it was already seen in Fig.~\ref{fig9}A.

The more interesting case is the nanohole limit ($H/R\rightarrow 0$ with constant $R$) where the curves seem to collapse onto a line, at least, for concentrations with Debye lengths smaller than $R$.
It is important that the slope of the line is $1$, so it can be given by the equation
\begin{equation}
 \lg \left(  \dfrac{\mathrm{Du}_{0}}{\mathrm{Du}_{\infty}} \right) = b +\lg\left( \dfrac{\lambda_{\mathrm{D}}}{H} \right) .
 \label{eq:linfit}
\end{equation} 
where $b$ is a fittable parameter. The resulting value is $b\approx 1.4$.
For the NP+LEMC simulations ($\mathrm{Du}_{\infty}=0.37$), the inflection point in the $H/R\rightarrow 0$ limit for $R=1$ nm and $U=200$ mV can be expressed as
\begin{equation}
 \mathrm{Du}_{0}=9.29 \frac{\lambda_{\mathrm{D}}}{H}.
\end{equation} 
Now if we want to bring the selectivity curves onto each other, namely, we want the scaling to work, we need to divide $\mathrm{Du}$ by this $\mathrm{Du}_{0}$ to define a new scaling parameter that we call the modified Dukhin number:
\begin{equation}
 \mathrm{mDu} = \mathrm{Du}\frac{1}{\beta}\frac{H}{\lambda_{\mathrm{D}}} = \frac{|\sigma| l_{\mathrm{B}}^{*}\lambda_{\mathrm{D}}H}{eR\beta},
 \label{eq:mDu}
\end{equation} 
where the $\beta$ parameter depends on $R$ and $U$. 
We have the most data for $U=200$ mV, so at this point, we can state the existence of the above scaling parameter only in the large voltage limit. 

\begin{figure*}[t!]
\begin{center}
\includegraphics*[width=0.95\textwidth]{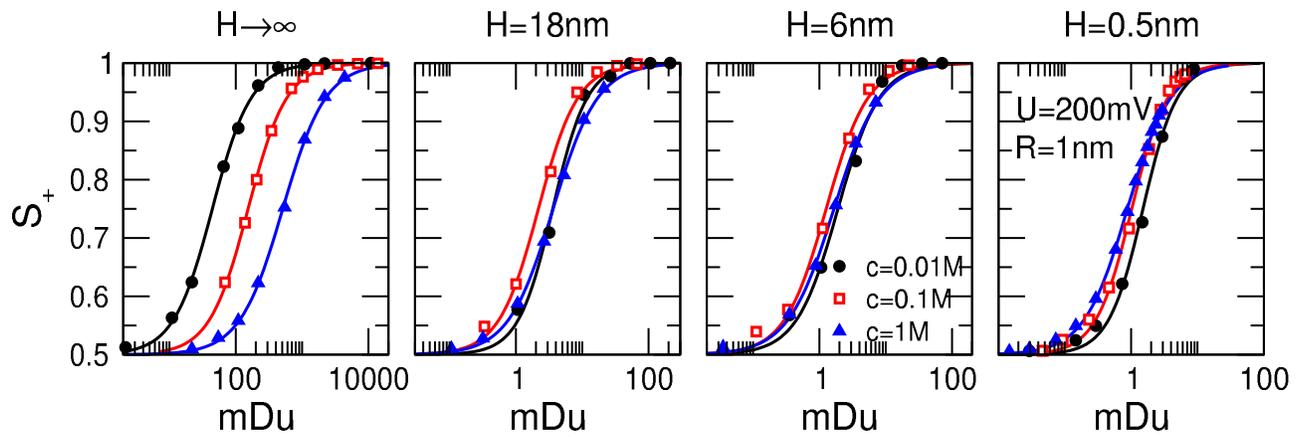}
\end{center}
\caption{Selectivity curves obtained by scanning the surface charge, $\sigma$, for nanopores of decreasing lengths from left to right (from $H\rightarrow\infty$ to $H=0.5$ nm) as functions of the modified Dukhin number (Eqs.~\ref{eq:mDu}). 
The radius of the pore and voltage are fixed ($R=1$ nm and $U=200$ mV).
The symbols have been obtained from NP+LEMC simulations (GCMC for $H\rightarrow\infty$) for different concentrations, while the lines are fitted sigmoid curves (Eq.~\ref{eq:sigmoid}).
}
\label{fig10}
\end{figure*}

The selectivity curves as functions of this modified Dukhin number are shown in Fig.~\ref{fig10} for the same cases included in Fig.~\ref{fig8}.
It is seen that the new scaling parameter works quite well for short pores: it brings the curves for different concentrations together and also the curves for different pore lengths because the inflection points are close to $1$.
As $H$ is increased ($H =18$ nm and $H\rightarrow \infty$), the curves become more separated and the inflection points are shifted.

The advantage of the mDu parameter is that it contains $H$ so it can account for the length of the pore.
Unfortunately, it is not appropriate for the long pores ($H\rightarrow \infty$).
The question arises whether we can construct a scaling parameter that is appropriate for both long and short pores.
At this point of our research, we have only a heuristic formula for the scaling parameter (let us denote it with $\mathrm{Du}'$) that is based on the two limits:
\begin{equation}
 \mathrm{Du}=\lim_{H\rightarrow \infty} \mathrm{Du}' \quad \quad \mathrm{and} \quad \quad \mathrm{mDu}=\lim_{H\rightarrow 0} \mathrm{Du}' .
\end{equation} 
This implies that $\mathrm{Du}'$ can be written as a mix of the two limiting cases.
The parameter that tuned how strongly the DLs at the entrances influence the pore interior is $\lambda_{\mathrm{D}}/H$, so we propose the following equation for the ``universal'' scaling parameter:
\begin{equation}
 \mathrm{Du}' = \mathrm{Du} \dfrac{1}{1+a(c) \dfrac{\lambda_{\mathrm{D}}}{H}}.
 \label{eq:newDu}
\end{equation} 
If $H$ is large, the $a(c)\lambda_{\mathrm{D}}/H$ term can be neglected next to $1$, so we recover $\mathrm{Du}$.
If $H$ is small, $1$ can be neglected next to $a(c)\lambda_{\mathrm{D}}/H$, so we recover $\mathrm{mDu}$.
The adjustable $a(c)$ parameter depends on the concentration, although we found it independent of concentration in the case of PNP.

In any case, by plotting $S_{+}$ against $\mathrm{Du}'$ we obtain a scaling that works for every value of $H$ investigated (see Fig.~3 of the SI).
This statement is valid rigorously only for $U=200$ mV and $R=1$ nm because that is the case that we investigated in detail. 
We are confident that $\mathrm{Du}'$ can be mixed for other values of $R$ as well, but we are not so sure about voltage. 

\begin{figure}[t!]
\begin{center}
\includegraphics*[width=0.35\textwidth]{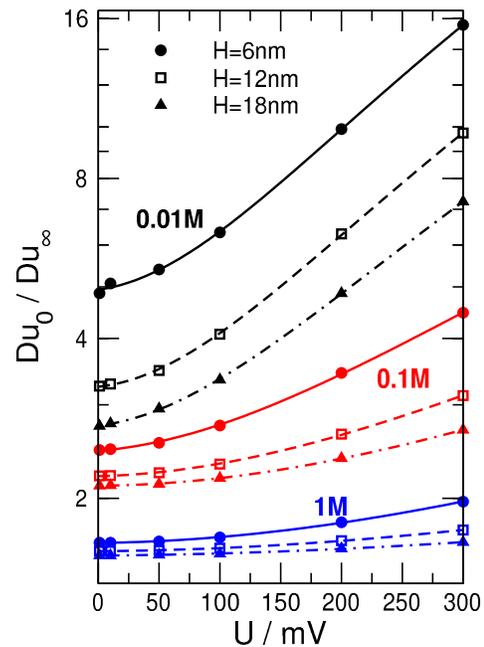}
\end{center}
\caption{The inflection point of the $S_{+}(\mathrm{Du})$ curves, $\mathrm{Du}_{0}$, normalized by $\mathrm{Du}_{\infty}$ as a function of voltage, $U$, as obtained from PNP calculations for different pore lengths, $H$, and concentrations, $c$ (pore radius is fixed at $R=1$ nm). The lines are quadratic fits.
}
\label{fig11}
\end{figure}

Voltage dependence has been studied with the PNP theory. 
The results are shown in Fig.~\ref{fig11}.
The general trend is that $\mathrm{Du}_{0}/\mathrm{Du}_{\infty}$ depends on $U^{2}$: the lines in Fig.~\ref{fig11} are quadratic fits.
The dependence on voltage is stronger at small concentrations, when the ionic diffuse layer is extended over a large space influencing the behavior inside the pore and voltage can distort this more diffuse DL more efficiently.
Phrasing in a different way, voltage produces more extended positive/negative DLs (shown in in Fig.~\ref{fig3}B) for dilute electrolytes.  

Also, the effect of voltage is stronger if the pore is shorter. 
This is because the charge in the DLs is larger in the case of the shorter pore (see Fig.~\ref{fig3}A and the Appendix for more detailed discussion) and the DLs can extend into the pore more deeply and can modify the ionic distributions. 

Summarized, the interplay of the various parameters of the system ($U$, $R$, $H$, $c$, and $\sigma$) is so intricate that it is difficult to come up with a ``universal'' scaling parameter. 
However, we can draw conclusions for the behavior of the nanopore for other combinations of the parameters on the basis of knowledge about the behavior for a specific parameter set.

\begin{figure*}[t!]
\begin{center}
\includegraphics*[width=0.6\textwidth]{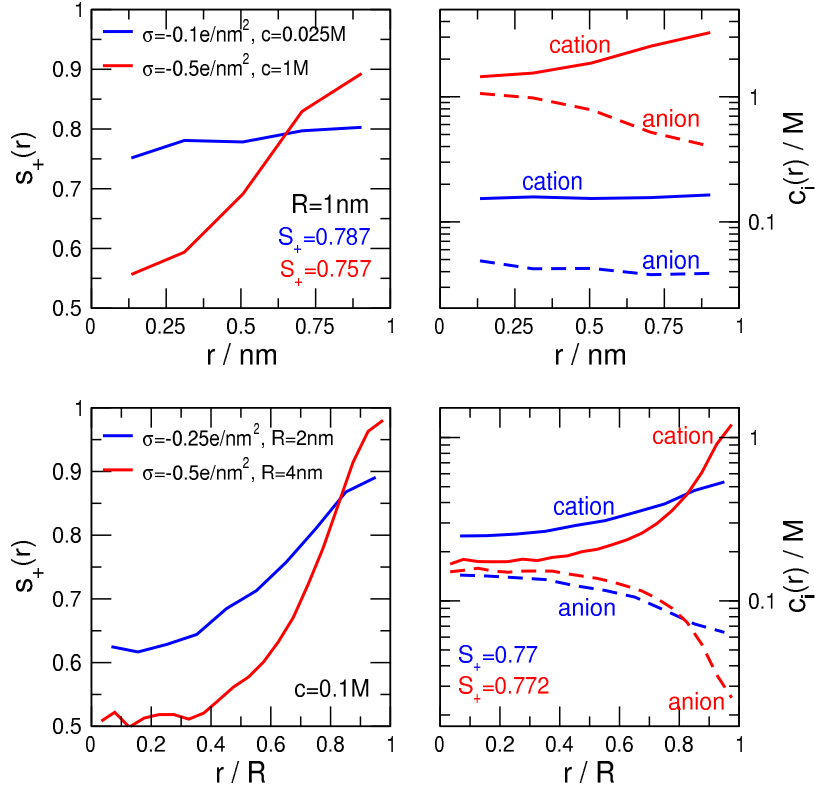}
\end{center}
\caption{Radial selectivity profiles ($s_{+}(r)$, left panels) and radial concentration profiles ($c_{i}(r)$, right panels) as obtained from NP+LEMC simulations for parameters indicated in the figure ($H=6$ nm and $U=200$ mV).
Top row: the pore radius is fixed ($R=1$ nm), while $\sigma$ and $c$ are changed in a way that the selectivity is similar for the two sets of curves (red and blue).
Bottom row: the concentration is fixed ($c=0.1$ M), while $\sigma$ and $R$ are changed in a way that the selectivity is similar for the two sets of curves (red and blue).
In the right panels, solid and dashed lines refer to cations and anions, respectively.
}
\label{fig6o}
\end{figure*}

\subsection{Volume and surface conduction vs.\ scaling} 

The Dukhin number was originally introduced as a parameter characterizing the ratio of the surface and volume conductions.
This works well in the limiting cases when either volume conduction or surface conduction dominates, e.g., for $\mathrm{Du}\ll 1$ or $\mathrm{Du}\gg 1$, or, equivalently, for $S_{+}\approx 0.5$ or $S_{+}\approx 1$.
In this work, however, we are interested in the domain in between, when the nanopore may contain both a region where rather the surface conduction dominates (near the wall) and a region where rather volume conduction dominates (in the middle of the pore around the centerline).

In this domain (around the inflection point), it is not obvious from the ``global'' selectivity value ($S_{+}$) or the Dukhin number alone to what degree surface and volume conductions are present in the two respective regions of the pore (close and far from the surface).
For example, if we have a large selectivity value $S_{+}=0.85$, it does not necessarily mean that there is no volume conduction in the pore.
Similarly, if we have a mildly selective value $S_{+}=0.65$, it does not necessarily mean that there is no surface conduction.

If we want more information, we need to look into the ``black box'' and investigate concentration and selectivity profiles as we already did in Fig.~\ref{fig6} for the nanotube limit.
In Fig.~\ref{fig6}, we already observed that the same selectivity can be achieved in different ways depending on the relation of $\lambda_{\mathrm{D}}/R$ and $\sigma$.
Different combinations of these parameters can produce the same Du, and, consequently, the same $S_{+}$, but the corresponding profiles behave differently.
Larger surface charge enhances separation of the cation and anion profiles near the wall, while large $\lambda_{\mathrm{D}}/R$ value enhances DL overlap in the centerline of the pore.

Figure \ref{fig6o} shows similar results for a finite pore ($H=6$ nm).
The top row shows selectivity and concentration profiles for two simulations that, due to scaling, produce about the same ``global'' selectivity, $S_{+}=0.787$ (blue) and $0.757$ (red). 
The blue curve is for a smaller surface charge ($\sigma =-0.1$ $e$/nm$^{2}$) and a smaller concentration ($c=0.025$ M), while the red curve is for a larger surface charge ($\sigma =-0.5$ $e$/nm$^{2}$) and a larger concentration ($c=1$ M) for fixed $R$.

The $s_{+}(r)$ curves (left panel) show that approximately the  same $S_{+}$ value can be achieved in two ways. 
In the case of the blue curve, the DL overlaps (small $c$, large $\lambda_{\mathrm{D}}$) so the coion (anion) is excluded over the whole cross section and selectivity is at a high level even if $\sigma$ is relatively small.
In the case of the red curve, the DL overlaps less (large $c$, small $\lambda_{\mathrm{D}}$) so a less selective bulk region is formed in the middle, but the larger surface charge ($\sigma=-0.5$ $e$/nm$^{2}$) produces a large counterion-coion separation near the wall and pulls up selectivity there.

The bottom panels show results for two simulations that produce approximately the same ``global'' selectivities, $S_{+}=0.77$ (blue) and $0.772$ (red) for the same concentration ($c=0.1$ M).
The blue curve is for a smaller surface charge ($\sigma =-0.25$ $e$/nm$^{2}$) and a smaller pore radius ($R=2$ nm), while the red curve is for a larger surface charge ($\sigma =-0.5$ $e$/nm$^{2}$) and a larger pore radius ($R=4$ nm).

The $s_{+}(r)$ curves (left panel) show similar behavior to those in the top panel.
In the case of the blue curve, the DL overlaps (small $R$) so there is less space for the formation of a bulk region in the middle.
Therefore, the selectivity is uniformly large even if $\sigma$ is relatively small.
In the case of the red curve, the degree of overlap of the DL is smaller (large $R$) so a less selective bulk region is formed in the middle, but the larger surface charge ($\sigma=-0.5$ $e$/nm$^{2}$) produces a large counterion-coion separation near the wall and pulls up selectivity there.

Looking into the ``black box'', however, is usually not possible in experiments.
Therefore, for a rule of thumb to decide whether volume or surface conduction dominates in a state point, we propose using the inflection point.

The inflection point offers itself as a transition point separating ``rather non-selective'' and ``rather selective'' regions.
The  derivative of the $S_{+}$ curve as a function of $\lg(\mathrm{Du})$ (or any suitable scaling parameter) is maximal in the inflection point where small changes in Du lead to relatively large changes in selectivity.

\section{Summary}

Scaling is a useful property of nanodevices, because we can estimate the device function from prior knowledge on its behavior for some well-measurable combination of the input parameters. 
Or, vice versa, we can estimate a missing not-so-well-measurable input parameter (pore charge, for example) from measurements for the selectivity.
Therefore, scaling may help us in solving inverse problems.

Starting from our previous work,~\cite{fertig_jpcc_2019} where we fixed the surface charge and the pore length for a bipolar nanopore (the device function is the rectification), we started this study with the ambition to include additional input parameters in the scaling parameter such as $\sigma$.
The parameters $c$, $R$, $\sigma$ have effects in the radial dimension of the nanopore.
They tune the degree of charge separation in the DL near the pore wall and the degree of overlap of the DLs.
These three parameters compose into the Dukhin number that turns out to be an appropriate scaling parameter for the infinitely long pore (the nanotube limit).

Including $H$ and $U$ in the parameter set, however, we switch on axial effects that proved to be extremely important for short pores.
The DLs that appear at the entrances of the nanopore near the membranes on the two sides change the electric field inside the pore, and, thus, the behavior of the ions.
Consequently, they change selectivity.

The interplay of the radial and axial effects, however, is not trivial and in most cases they are hard to separate.
Therefore, finding a ``universal'' scaling parameter turned out to be a difficult if not insurmountable task. 
In this work, we followed the strategy of fixing some parameters and studying the effect of others, and, focusing on the limiting cases (nanotube and nanohole).
With this procedure, we were able to describe the dependence of scaling in term of the shift of the inflection point on the respective parameters.

In the nanohole limit, at least, for large voltage, we proposed a modified Dukhin number that includes $H$ and depends on $\lambda_{\mathrm{D}}$ instead of $\lambda_{\mathrm{D}}^{2}$, that is, it depends on $\sqrt{c}$ instead of $c$.

The inflection point of the $S_{+}(\mathrm{Du})$ curve has a special role in our treatment.
It may characterize the transition point that separates the ``rather selective'' and the ``rather non-selective'' state points.

For electrolytes containing multivalent ions (2:2, 2:1, and 3:1, for example) interesting phenomena beyond the mean-field treatment may occur due to strong ionic correlations such as overcharging and charge inversion.~\cite{fertig_pccp_2020}
Ionic correlations may also be enhanced by using a solvent of smaller dielectric constant. 
These will be reported in subsequent publications.

\section*{Acknowledgements}
\label{sec:ack}

We gratefully acknowledge  the financial support of the National Research, Development and Innovation Office -- NKFIH K124353. 
Present article was published in the frame of the project GINOP-2.3.2-15-2016-00053 (``Development of engine fuels with high hydrogen content in their molecular structures (contribution to sustainable mobility)'').
We are grateful to Dirk Gillespie and Tam\'as Krist\'of for inspiring discussions. 
We are grateful to the reviewers for their inspiring remarks that led to a substantially reworked and improved manuscript.


\appendix
\section{Computational methods}
\label{sec:me4hods}

In this work, we use two procedures that have the common denominator that both apply the NP transport equation~\cite{nernst_zpc_1888,planck_apc_1890} to compute the ionic flux. 
The difference is that the two procedures use different methods to relate $c_{i}(\mathbf{r})$ to $\mu_{i}(\mathbf{r})$.

The LEMC method~\cite{boda_jctc_2012} is a particle simulation technique devised for a non-equilibrium situation, where $\mu_{i}(\mathbf{r})$ is not constant globally.
We divide the simulation into small volume elements, $V^{\alpha}$, and assume local thermodynamic equilibrium in each, namely, we assign $\mu_{i}^{\alpha}$ values to each.
Then, we perform particle displacements and particle insertions/deletions with the same acceptance criteria as we do in a GCMC simulation, but with the volume, particle number ($N_{i}^{\alpha}$), and chemical potential of the volume element in which we peform an MC step.
Therefore, the LEMC technique is an adaptation of the GCMC technique for a system that is not at equilibrium globally.

The resulting method, coined NP+LEMC, solves the problem iteratively on the basis of the scheme
\begin{equation}
\mu^{\alpha}_{i}[n] \, \xrightarrow{\mathrm{LEMC}}  \,  c^{\alpha}_{i}[n] \,  \xrightarrow{\mathrm{NP}} \,  \mathbf{j}^{\alpha}_{i}[n] \, 
\xrightarrow{\nabla \cdot \mathbf{j}=0} 
\,\, \mu^{\alpha}_{i}[n+1] ,
\label{eq:circle}
\end{equation} 
where $c_{i}^{\alpha}[n]$ is the concentration in volume element $V^{\alpha}$ obtained from an LEMC simulation in the $n$th iteration.
The chemical potential for the next $[n{+}1]$th iteration is obtained by assuming that the flux $\mathbf{j}_{i}^{\alpha}$ computed from it and from $c_{i}^{\alpha}[n]$ satisfies the continuity equation, $\nabla \cdot \mathbf{j}_{i}=0$.
Details are found in Refs.~\onlinecite{boda_jctc_2012,boda_jml_2014,boda_arcc_2014,fertig_hjic_2017}.

The importance of the LEMC technique is that it can take into account the correlations between ions beyond the mean-field approximation including the finite size of ions.

In the PNP theory, we relate $c_{i}(\mathbf{r})$ to $\mu_{i}(\mathbf{r})$ via the Poisson-Boltzmann (PB) theory where the ions are modeled as point charges interacting with the average electrical potential, $\Phi(\mathbf{r})$,  exerted by all the ions in the system.
In this mean-field approach, the electrolyte is assumed to be an ideal solution with the electrochemical potential
\begin{equation*}
 \mu_{i}(\mathbf{r}) = \mu_{i}^{0} + kT\ln c_{i}(\mathbf{r}) + z_{i}e\Phi(\mathbf{r}),
\end{equation*}
where $\mu_{i}^{0}$ is a reference chemical potential independent of the location.
Note that an excess chemical potential $\mu_{i}^{\mathrm{ex}}(\mathbf{r})$, is added to this expression when ionic correlations are taken into account as they are in the LEMC method.
Poisson's equation and the continuity equation are also satisfied in the solution of the PNP theory.

Here, we solve the system with the Scharfetter–Gummel scheme.~\cite{gummel1964self} 
A 2D finite element method is used with $20-60$ thousand elements in a triangular mesh.
Details are found in Ref.~\onlinecite{matejczyk_jcp_2017}.

The constant $\sigma$ surface charge is assured via a Neumann boundary condition in PNP, while fractional point charges are placed on a rectangular grid of width $0.2$ nm in LEMC.

In both methods, appropriate boundary conditions were applied in the two baths at the wall of the cylinder that confines the finite simulation cell. 
Dirichlet boundary conditions were applied for $\Phi(\mathrm{r})$; the difference of the applied potential on the two sides of the membrane specifies the applied voltage, $U$.

\begin{figure}[t!]
\begin{center}
\includegraphics*[width=0.35\textwidth]{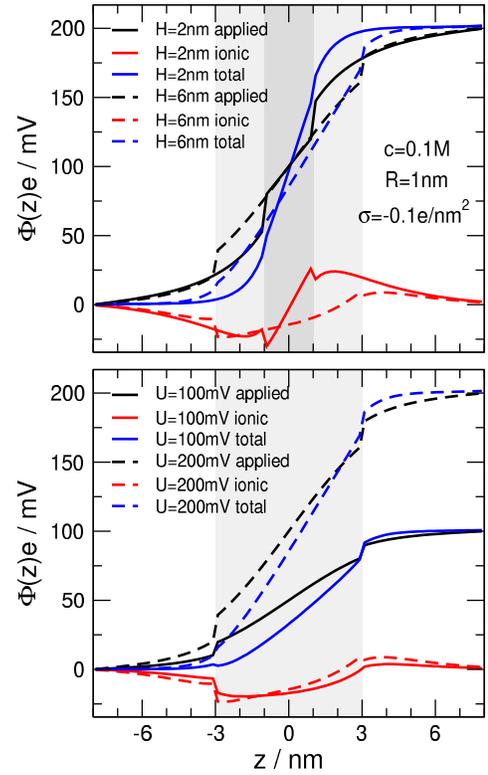}
\end{center}
\caption{Axial mean potential profiles as provided by NP+LEMC simulations.
In addition to the total, we show the applied electrical potential produced by the electrodes represented by the Dirichlet boundary conditions and the ``ionic term'' produces by all the charges in the system (ions and pore charges).
The top panel shows the results for two values of $H$ for $U=200$ mV, while the bottom panel shows the results for two values of $U$ for $H=6$ nm.
}
\label{figapp}
\end{figure}

In the main text, we show that DLs with opposite signs are formed at the entrances of the pore on the two sides of the membrane. 
This ``dipole field'' exerts a counterfield against the applied voltage that is computed by solving Laplace's equation with the Dirichlet boundary conditions on the two half-cylinders on the two sides of the membrane ($\Phi=0$ on the left hand side, while $\Phi=U$ on the right hand side).
The ionic counterfield is necessary because the total mean electrical potential must be close to horizontal in the baths with small slopes due to the condition that the resistance of the baths is small.
This is shown in Fig.~\ref{figapp} for two different pore lengths (top panel) and voltages (bottom panel). 

If we denote the degree of charge separation in the DLs with $Q$, we can write that $U\sim HQ$; the ``dipole field'' counterbalances the applied field.
A larger voltage then creates a stronger charge separation: larger $U$ requires larger $Q$.
A shorter pore, $H$, for the same $U$ also requires larger $Q$ to obtain the same dipole field in the baths.

Bath concentrations were assumed to be the same on the two sides of the membrane, $c$, though the methodology would be able to handle asymmetrical systems as well. 
In the 1:1 electrolyte considered here $c_{+}=c_-=c$ in both baths.

For the diffusion coefficient profile, $D_{i}(\mathbf{r})$, we use a piecewise constant function, where the value in the baths is $1.334\times 10^{-9}$ m$^{2}$s$^{-1}$ for both ionic species, while it is the tenth of that inside the pore, $D_{i}^{\mathrm{pore}}$, as in our earlier works.~\cite{matejczyk_jcp_2017,madai_jcp_2017,madai_pccp_2018,fertig_jpcc_2019,fertig_pccp_2020}
These particular choices do not qualitatively affect our conclusions.

In this model calculation, where $R_{+}=R_-$ and $D_{+}(\mathbf{r})=D_-(\mathbf{r})$ for a 1:1 electrolyte, so $S_{+}=0.5$ exactly for a perfectly non-selective pore ($\sigma=0$, for example).


\begin{thebibliography}{10}

\bibitem{fertig_jpcc_2019}
D.~Fertig, B.~Matejczyk, M.~Valisk{\'{o}}, D.~Gillespie, and D.~Boda.
\newblock Scaling behavior of bipolar nanopore rectification with multivalent
  ions.
\newblock {\em J. Phys. Chem. C}, 123(47):28985--28996, 2019.

\bibitem{blum_mp_1975}
L.~Blum.
\newblock Mean spherical model for asymmetric electrolytes.
\newblock {\em Mol. Phys.}, 30(5):1529--1535, 1975.

\bibitem{blum_jcp_1977}
L.~Blum and J.~S. Hoeye.
\newblock Mean spherical model for asymmetric electrolytes. 2. {Thermodynamic}
  properties and the pair correlation function.
\newblock {\em J. Phys. Chem.}, 81(13):1311--1316, 1977.

\bibitem{nonner_bj_2000}
W.~Nonner, L.~Catacuzzeno, and B.~Eisenberg.
\newblock {Binding and Selectivity in {L}-Type Calcium Channels: {A} Mean
  Spherical Approximation.}
\newblock {\em Biophys. J.}, 79(4):1976--1992, 2000.

\bibitem{madai_pccp_2018}
E.~M\'adai, B.~Matejczyk, A.~Dallos, M.~Valisk\'o, and D.~Boda.
\newblock Controlling ion transport through nanopores: modeling transistor
  behavior.
\newblock {\em Phys. Chem. Chem. Phys.}, 20(37):24156--24167, 2018.

\bibitem{Siwy_2004}
Z.~Siwy, E.~Heins, C.~C. Harrell, P.~Kohli, and C.~R. Martin.
\newblock Conical-nanotube ion-current rectifiers:~ the role of surface charge.
\newblock {\em J. Am. Chem. Soc.}, 126(35):10850--10851, 2004.

\bibitem{garaj_n_2010}
S.~Garaj, W.~Hubbard, A.~Reina, J.~Kong, D.~Branton, and J.~A. Golovchenko.
\newblock Graphene as a subnanometre trans-electrode membrane.
\newblock {\em Nature}, 467(7312):190--193, 2010.

\bibitem{garaj_pnas_2013}
S.~Garaj, S.~Liu, J.~A. Golovchenko, and D.~Branton.
\newblock Molecule-hugging graphene nanopores.
\newblock {\em Proc. Nat. Acad. Sci.}, 110(30):12192--12196, 2013.

\bibitem{nernst_zpc_1888}
W.~Nernst.
\newblock Zur kinetik der in l\"{o}sung befindlichen k\"{o}rper.
\newblock {\em Zeitschrift für Physikalische Chemie}, 2(1), 1888.

\bibitem{planck_apc_1890}
M.~Planck.
\newblock Ueber die erregung von electricit\"{a}t und w\"{a}rme in
  electrolyten.
\newblock {\em Annalen der Physik und Chemie}, 275(2):161--186, 1890.

\bibitem{malasics_jcp_2008}
A.~Malasics, D.~Gillespie, and D.~Boda.
\newblock Simulating prescribed particle densities in the grand canonical
  ensemble using iterative algorithms.
\newblock {\em J. Chem. Phys.}, 128(12):124102, 2008.

\bibitem{malasics_jcp_2010}
A.~Malasics and D.~Boda.
\newblock An efficient iterative grand canonical monte carlo algorithm to
  determine individual ionic chemical potentials in electrolytes.
\newblock {\em J. Chem. Phys.}, 132(24):244103, 2010.

\bibitem{levine_jcis_1975}
S.~Levine, J.~R. Marriott, G.~Neale, and N.~Epstein.
\newblock Theory of electrokinetic flow in fine cylindrical capillaries at high
  zeta-potentials.
\newblock {\em J. Coll. Interf. Sci.}, 52(1):136--149, 1975.

\bibitem{balme_sr_2015}
S.~Balme, F.~Picaud, M.~Manghi, J.~Palmeri, M.~Bechelany, S.~Cabello-Aguilar,
  A.~Abou-Chaaya, P.~Miele, E.~Balanzat, and J.~M. Janot.
\newblock Ionic transport through sub-10~nm diameter hydrophobic high-aspect
  ratio nanopores: experiment, theory and simulation.
\newblock {\em Sci. Rep.}, 5(1), 2015.

\bibitem{uematsu_jpcb_2018}
Y.~Uematsu, R.~R. Netz, L.~Bocquet, and D.~J. Bonthuis.
\newblock Crossover of the power-law exponent for carbon nanotube conductivity
  as a function of salinity.
\newblock {\em J. Phys. Chem. B}, 122(11):2992--2997, 2018.

\bibitem{green_jcp_2021}
Y.~Green.
\newblock Ion transport in nanopores with highly overlapping electric double
  layers.
\newblock {\em J. Chem. Phys.}, 154(8):084705, 2021.

\bibitem{bazant_pre_2004}
M.~Z. Bazant, K.~Thornton, and A.~Ajdari.
\newblock Diffuse-charge dynamics in electrochemical systems.
\newblock {\em Phys. Rev. E}, 70(2):021506, 2004.

\bibitem{chu_pre_2006}
K.~T. Chu and M.~Z. Bazant.
\newblock Nonlinear electrochemical relaxation around conductors.
\newblock {\em Phys. Rev. E}, 74(1):011501, 2006.

\bibitem{bocquet_chemsocrev_2010}
L.~Bocquet and E.~Charlaix.
\newblock Nanofluidics, from bulk to interfaces.
\newblock {\em Chem. Soc. Rev.}, 39(3):1073--1095, 2010.

\bibitem{bikerman_1940}
J.~J. Bikerman.
\newblock Electrokinetic equations and surface conductance. a survey of the
  diffuse double layer theory of colloidal solutions.
\newblock {\em Trans. Farad. Soc.}, 35:154, 1940.

\bibitem{dukhin_advcollsci_1993}
S.S. Dukhin.
\newblock Non-equilibrium electric surface phenomena.
\newblock {\em Adv. Coll. Interf. Sci.}, 44:1--134, 1993.

\bibitem{lyklema_book_1995}
J.~J. Lyklema, A.~de~Keizer, B.H. Bijsterbosch, G.J. Fleer, and M.A.
  Cohen~Stuart (Eds.).
\newblock {\em Solid-Liquid Interfaces}.
\newblock Fundamentals of Interface and Colloid Science 2. Elsevier, Academic
  Press, 1995.

\bibitem{khair_jfm_2008}
A.~S. Khair and T.~M. Squires.
\newblock Surprising consequences of ion conservation in electro-osmosis over a
  surface charge discontinuity.
\newblock {\em J. Fluid Mech.}, 615:323--334, 2008.

\bibitem{das_langmuir_2010}
S.~Das and S.~Chakraborty.
\newblock Effect of conductivity variations within the electric double layer on
  the streaming potential estimation in narrow fluidic confinements.
\newblock {\em Langmuir}, 26(13):11589--11596, 2010.

\bibitem{zangle_csr_2010}
T.~A. Zangle, A.~Mani, and J.~G. Santiago.
\newblock Theory and experiments of concentration polarization and ion focusing
  at microchannel and nanochannel interfaces.
\newblock {\em Chem. Soc. Rev.}, 39(3):1014, 2010.

\bibitem{lee_nanolett_2012}
C.~Lee, L.~Joly, A.~Siria, A.-L. Biance, R.~Fulcrand, and L.~Bocquet.
\newblock Large apparent electric size of solid-state nanopores due to
  spatially extended surface conduction.
\newblock {\em Nano Lett.}, 12(8):4037--4044, 2012.

\bibitem{yeh_ijc_2014}
H.-C. Yeh, M.~Wang, C.-C. Chang, and R.-J. Yang.
\newblock Fundamentals and modeling of electrokinetic transport in
  nanochannels.
\newblock {\em Israel J. Chem.}, 54(11-12):1533--1555, 2014.

\bibitem{ma_acssens_2017}
Y.~Ma, J.~Guo, L.~Jia, and Y.~Xie.
\newblock Entrance effects induced rectified ionic transport in a
  nanopore/channel.
\newblock {\em {ACS} Sensors}, 3(1):167--173, 2017.

\bibitem{xiong_scc_2019}
T.~Xiong, K.~Zhang, Y.~Jiang, P.~Yu, and L.~Mao.
\newblock Ion current rectification: from nanoscale to microscale.
\newblock {\em Sci. China Chem.}, 62(10):1346--1359, 2019.

\bibitem{poggioli_jpcb_2019}
A.~R. Poggioli, A.~Siria, and L.~Bocquet.
\newblock Beyond the tradeoff: Dynamic selectivity in ionic transport and
  current rectification.
\newblock {\em J. Phys. Chem. B}, 123(5):1171--1185, 2019.

\bibitem{dalcengio_jcp_2019}
S.~Dal Cengio and I.~Pagonabarraga.
\newblock Confinement-controlled rectification in a geometric nanofluidic
  diode.
\newblock {\em J. Chem. Phys.}, 151(4):044707, 2019.

\bibitem{kavokine_annualrev_2020}
N.~Kavokine, R.~R. Netz, and L.~Bocquet.
\newblock Fluids at the nanoscale: From continuum to subcontinuum transport.
\newblock {\em Annu. Rev. Fluid Mech.}, 53(1), 2020.

\bibitem{noh_acsnano_2020}
Y.~Noh and N.~R. Aluru.
\newblock Ion transport in electrically imperfect nanopores.
\newblock {\em {ACS} Nano}, 14(8):10518--10526, 2020.

\bibitem{levy_jcis_2020}
A.~Levy, J.~P. de~Souza, and M.~Z. Bazant.
\newblock Breakdown of electroneutrality in nanopores.
\newblock {\em J. Coll. Inter. Sci.}, 579:162--176, 2020.

\bibitem{fertig_pccp_2020}
D.~Fertig, M.~Valisk{\'{o}}, and D.~Boda.
\newblock Rectification of bipolar nanopores in multivalent electrolytes:
  effect of charge inversion and strong ionic correlations.
\newblock {\em Phys. Chem. Chem. Phys.}, 22(34):19033--19045, 2020.

\bibitem{boda_jctc_2012}
D.~Boda and D.~Gillespie.
\newblock Steady state electrodiffusion from the {Nernst-Planck} equation
  coupled to {Local Equilibrium Monte Carlo} simulations.
\newblock {\em J. Chem. Theor. Comput.}, 8(3):824--829, 2012.

\bibitem{boda_jml_2014}
D.~Boda, R.~Kov\'acs, D.~Gillespie, and T.~Krist\'of.
\newblock Selective transport through a model calcium channel studied by
  {Local} {Equilibrium} {Monte} {Carlo} simulations coupled to the
  {Nernst}-{Planck} equation.
\newblock {\em J. Mol. Liq.}, 189:100--112, 2014.

\bibitem{boda_arcc_2014}
D.~Boda.
\newblock In R.~A. Wheeler, editor, {\em Ann. Rep. Comp. Chem.}, volume~10,
  chapter 5 {Monte Carlo} Simulation of Electrolyte Solutions in Biology: {In}
  and Out of Equilibrium, pages 127--163. Elsevier, 2014.

\bibitem{fertig_hjic_2017}
D.~Fertig, E.~M\'adai, M.~Valisk\'o, and D.~Boda.
\newblock Simulating ion transport with the {NP+LEMC} method. {Applications} to
  ion channels and nanopores.
\newblock {\em Hung. J. Ind. Chem.}, 45(1):73--84, 2017.

\bibitem{gummel1964self}
H.~K. Gummel.
\newblock A self-consistent iterative scheme for one-dimensional steady state
  transistor calculations.
\newblock {\em IEEE Transactions on electron devices}, 11(10):455--465, 1964.

\bibitem{matejczyk_jcp_2017}
B.~Matejczyk, M.~Valisk{\'{o}}, M.-T. Wolfram, J.-F. Pietschmann, and D.~Boda.
\newblock Multiscale modeling of a rectifying bipolar nanopore: {Comparing}
  {Poisson-Nernst-Planck} to {Monte Carlo}.
\newblock {\em J. Chem. Phys.}, 146(12):124125, 2017.

\bibitem{madai_jcp_2017}
E.~M\'adai, M.~Valisk\'o, A.~Dallos, and D.~Boda.
\newblock Simulation of a model nanopore sensor: {Ion} competition underlines
  device behavior.
\newblock {\em J. Chem. Phys.}, 147(24):244702, 2017.

\end{thebibliography}

\end{document}